\documentclass[prx, twocolumn, superscriptaddress, longbibliography]{revtex4-1}
\usepackage{bm, amsmath, amsfonts, amssymb, braket}
\usepackage[dvipdfmx]{graphicx}
\usepackage{float}
\usepackage{color}

\newcommand{\ii}{\text{i}}

\begin{document}

\title{Real spectra in non-Hermitian topological insulators}

\author{Kohei Kawabata}
	\email{kawabata@cat.phys.s.u-tokyo.ac.jp}
	\affiliation{Department of Physics, University of Tokyo, 7-3-1 Hongo, Bunkyo-ku, Tokyo 113-0033, Japan}
\author{Masatoshi Sato}
	\affiliation{Yukawa Institute for Theoretical Physics, Kyoto University, Kyoto 606-8502, Japan}

\date{\today}

\begin{abstract} 
Spectra of bulk or edges in topological insulators are often made complex by non-Hermiticity. Here, we show that symmetry protection enables entirely real spectra for both bulk and edges even in non-Hermitian topological insulators. In particular, we demonstrate entirely real spectra without non-Hermitian skin effects due to a combination of pseudo-Hermiticity and Kramers degeneracy. This protection relies on nonspatial fundamental symmetry and has stability against disorder. As an illustrative example, we investigate a non-Hermitian extension of the Bernevig-Hughes-Zhang model. The helical edge states exhibit oscillatory dynamics due to their nonorthogonality as a unique non-Hermitian feature.
\end{abstract}

\maketitle

\section{Introduction}

Physics of non-Hermitian systems has generated considerable recent research interest~\cite{Konotop-review, Christodoulides-review}. Non-Hermiticity appears, for example, in open classical~\cite{Makris-08, Klaiman-08, Guo-09, Ruter-10, Lin-11, Regensburger-12, Feng-13, Peng-14, Wiersig-14, Hodaei-17, Chen-17} and quantum~\cite{Brody-12, Lee-14, Li-19, KAU-17, Hamazaki-19, Xiao-19, Wu-19, Yamamoto-19, Naghiloo-19, Matsumoto-19} systems as a consequence of the external environment. Despite non-Hermiticity, Hamiltonians can have entirely real spectra if parity-time symmetry~\cite{Bender-98, *Bender-02} or pseudo-Hermiticity~\cite{Mostafazadeh-02-1, *Mostafazadeh-02-2, *Mostafazadeh-02-3} is respected. Disorder can also give rise to the real spectra in time-reversal-invariant non-Hermitian systems~\cite{Hatano-Nelson-96, *Hatano-Nelson-97}. The reality of the spectra ensures the stability of the systems even in the presence of non-Hermiticity. On the other hand, when non-Hermiticity is sufficiently strong, the symmetry is spontaneously broken and some eigenenergies form complex-conjugate pairs. An exceptional point appears between the two phases, at which the eigenstates coalesce with each other~\cite{Heiss-12}. The real spectra and exceptional points were experimentally observed in a number of classical and quantum systems, such as a photonic lattice~\cite{Regensburger-12}, a microcavity~\cite{Peng-14}, single photons~\cite{Xiao-19}, a nitrogen-vacancy center~\cite{Wu-19}, and superconducting qubits~\cite{Naghiloo-19}.

Much research in recent years has focused on topological characterization of non-Hermitian systems~\cite{Ota-review, Bergholtz-review} both in theory~\cite{Rudner-09, Sato-11, *Esaki-11, Hu-11, Schomerus-13, Malzard-15, Lee-16, Zeng-16, Leykam-17, Xu-17, Menke-17, Shen-18, *Kozii-17, Takata-18, KAKU-18, Gong-18, YW-18, *YSW-18, KHGAU-19, Kunst-18, KSU-18, McDonald-18, Lee-19, Budich-19, Okugawa-19, Liu-19, Yoshida-19, *Kimura-19, Zhou-19, Kunst-19, KSUS-19, ZL-19, Rui-19, KBS-19, Yokomizo-19, McClarty-19, Song-19-Lindblad, Bergholtz-19, Guo-19, Hockendorf-19, Herviou-19-ES, Chang-19, Zeng-20, Zhang-20} and in experiments~\cite{Poli-15, Zeuner-15, Zhen-15, Weimann-17, Xiao-17, St-Jean-17, Bahari-17, Harari-18, *Bandres-18, Cerjan-19, Zhao-19, Brandenbourger-19-skin-exp, *Ghatak-19-skin-exp, Helbig-19-skin-exp, *Hofmann-19-skin-exp, Xiao-19-skin-exp, Weidemann-20-skin-exp}. Non-Hermiticity alters the fundamental nature of the topological classification of phases of matter~\cite{Gong-18, KSUS-19, KBS-19} and the bulk-boundary correspondence~\cite{Lee-16, Kunst-18, YW-18, *YSW-18, Lee-19, Kunst-19, Yokomizo-19}. Furthermore, the interplay of non-Hermiticity and topology leads to unique phenomena and functionalities that have no counterparts in conventional systems. A prime example is topological lasers~\cite{St-Jean-17, Bahari-17, Harari-18, *Bandres-18, Zhao-19}. Because of the judicious designs, they possess the real spectra for the bulk but the complex spectra for the edges; whereas the bulk states remain stable, the edge states are amplified, resulting in high-efficiency lasers protected by topology. 

Despite the significance of the reality of spectra, Ref.~\cite{Hu-11}, which is one of the earliest works on non-Hermitian topological systems~\cite{Rudner-09, Esaki-11, Hu-11}, showed that entirely real spectra of both bulk and edges are impossible in a large class of non-Hermitian topological insulators with parity-time symmetry. For example, when we introduce balanced gain and loss to the Su-Schrieffer-Heeger model~\cite{SSH-79} without breaking chiral symmetry (pseudo-anti-Hermiticity), the bulk spectrum remains real, but a pair of zero-energy edge states acquires nonzero imaginary eigenenergies~\cite{Sato-11, *Esaki-11, Schomerus-13, Weimann-17, St-Jean-17}. On the other hand, when we introduce asymmetric hopping to the Su-Schrieffer-Heeger model~\cite{SSH-79} without breaking sublattice symmetry, the entirely real spectrum for both bulk and edges can be realized under the open boundary conditions~\cite{Lee-16, Kunst-18, YW-18}; however, it relies on the non-Hermitian skin effect and the spectrum becomes complex under the periodic boundary conditions. Remarkably, Ref.~\cite{Hu-11} assumes no symmetry other than parity-time symmetry and mentions possible exceptions of its theorem due to particle-hole or point-group symmetry. In fact, a $p$-wave topological superconducting wire with balanced gain and loss, which is described by a non-Hermitian extension of the Kitaev chain~\cite{Kitaev-01} with parity-time symmetry, can possess the entirely real spectrum even in the presence of Majorana edge states~\cite{Zeng-16, Menke-17, KAKU-18}. By contrast, non-Hermitian topological insulators with entirely real spectra have yet to be known. Although the reality of spectra is relevant to the stability of non-Hermitian systems, the real spectra in non-Hermitian topological insulators have still been elusive.

In this work, we show that symmetry protection enables the entirely real spectra for both bulk and edges even in non-Hermitian topological insulators. This protection is due to nonspatial symmetry and stable against disorder. In Sec.~\ref{sec: real spectra}, we demonstrate that generic time-reversal-invariant topological insulators in two dimensions can have real spectra even in the presence of non-Hermiticity as long as reciprocity (a variant of time-reversal symmetry in non-Hermitian systems) and pseudo-Hermiticity are respected. As shown in Sec.~\ref{sec: Dirac} with a continuum Dirac Hamiltonian, the discussions in Ref.~\cite{Hu-11} are not directly applicable because of additional pseudo-Hermiticity and reciprocity. As an illustrative example, we investigate a non-Hermitian extension of the Bernevig-Hughes-Zhang (BHZ) model~\cite{BHZ-06} in Sec.~\ref{sec: BHZ}. We explicitly show that it indeed has a real spectrum by both numerical and analytical calculations. Despite the real spectrum, it shows phenomena unique to non-Hermitian systems. In particular, the helical edge states exhibit oscillatory dynamics since they are nonorthogonal, as shown in Sec.~\ref{sec: power oscillation}. We conclude this work in Sec.~\ref{sec: conclusion}. In Appendix~\ref{appendix: BHZ-complex}, we investigate another non-Hermitian extension of the BHZ model that is protected by time-reversal symmetry and possesses the complex edge spectrum.

\section{Real spectra due to symmetry protection}
	\label{sec: real spectra}

\subsection{Symmetry and topology}

We begin with a generic Hermitian Hamiltonian $H \left( \bm{k} \right)$ in two dimensions that respects time-reversal symmetry:
\begin{equation}
\mathcal{T} H^{*} \left( \bm{k} \right) \mathcal{T}^{-1}
= H \left( - \bm{k} \right),\quad
\mathcal{T}\mathcal{T}^{*} = -1,
	\label{eq: TRS (H)}
\end{equation}
where $H \left( \bm{k} \right)$ is a Bloch Hamiltonian, and $\mathcal{T}$ is a unitary matrix (i.e., $\mathcal{T} \mathcal{T}^{\dag} = \mathcal{T}^{\dag}\mathcal{T} = 1$). The topological phase of $H \left( \bm{k} \right)$ is characterized by the $\mathbb{Z}_{2}$ invariant, which induces the quantum spin Hall effect accompanying helical edge states~\cite{Kane-Mele-05-QSH, *Kane-Mele-05-Z2, BHZ-06, Konig-07}. Moreover, we consider additional unitary symmetry:
\begin{equation}
\eta H \left( \bm{k} \right) \eta^{-1} = H \left( \bm{k} \right),\quad
\eta^{2} = 1,
	\label{eq: unitary (H)}
\end{equation}
where $\eta$ is a unitary and Hermitian matrix (i.e., $\eta \eta^{\dag} = \eta^{\dag} \eta = 1$). We assume that these symmetry anticommutes with each other:
\begin{equation}
\mathcal{T} \eta^{*}
= -\eta \mathcal{T}.
	\label{eq: TRS - unitary}
\end{equation}

For example, the BHZ model~\cite{BHZ-06} respects these symmetry in Eqs.~(\ref{eq: TRS (H)}), (\ref{eq: unitary (H)}), and (\ref{eq: TRS - unitary}) with $\mathcal{T} = \ii \sigma_{y}$ and $\eta = \sigma_{z}$:
\begin{eqnarray}
&&H_{\rm BHZ} \left( {\bm k} \right)
= \left( m + t \cos k_{x} + t \cos k_{y} \right) \tau_{z} \nonumber \\
&&\qquad\qquad\qquad\quad + t \left( \sin k_{y} \right) \tau_{y} + t \left( \sin k_{x} \right) \sigma_{z} \tau_{x}.
	\label{eq: H-BHZ}
\end{eqnarray}
Here, Pauli matrices $\sigma_{i}$'s and $\tau_{i}$'s ($i = x, y, z$) describe the spin and orbital degrees of freedom, respectively. The BHZ model describes mercury telluride-cadmium telluride semiconductor quantum wells that host the quantum spin Hall effect, in which the unitary symmetry in Eq.~(\ref{eq: unitary (H)}) represents the conservation of spin.

As a non-Hermitian generalization of these symmetry, we consider a generic non-Hermitian Hamiltonian $H \left( \bm{k} \right)$ in two dimensions that respects
\begin{eqnarray}
\mathcal{T} H^{T} \left( \bm{k} \right) \mathcal{T}^{-1}
&=& H \left( - \bm{k} \right),\quad
\mathcal{T}\mathcal{T}^{*} = -1,
	\label{eq: TRS} \\
\eta H^{\dag} \left( \bm{k} \right) \eta^{-1} 
&=& H \left( \bm{k} \right),\quad
\eta^{2} = 1,
	\label{eq: pH}
\end{eqnarray}
where unitary matrices $\mathcal{T}$ and $\eta$ anticommute with each other [Eq.~(\ref{eq: TRS - unitary})]. Here, Eqs.~(\ref{eq: TRS}) and (\ref{eq: pH}) reduce to Eqs.~(\ref{eq: TRS (H)}) and (\ref{eq: unitary (H)}) in the presence of Hermiticity [i.e., $H^{\dag} \left( \bm{k} \right) = H \left( \bm{k} \right)$], respectively. When Eq.~(\ref{eq: TRS}) is satisfied, the scattering matrix $S$ respects $\mathcal{T} S^{T} \mathcal{T}^{-1} = S$, and hence the scattering processes are reciprocal~\cite{Beenakker-97, *Beenakker-15}. For example, an incoming spin-up wave is related to an outgoing spin-down wave because of $\mathcal{T} S^{T} \mathcal{T}^{-1} = S$.  Consequently, Eq.~(\ref{eq: TRS}) describes reciprocity in non-Hermitian systems and is relevant, for example, in mesoscopic systems~\cite{Beenakker-97, *Beenakker-15} and open quantum systems~\cite{Hamazaki-19-RMT, Lieu-19, Sa-19}. It is also notable that this symmetry is a variant of time-reversal symmetry and called ``TRS$^{\dag}$" in Ref.~\cite{KSUS-19}. On the other hand, Eq.~(\ref{eq: pH}) denotes pseudo-Hermiticity~\cite{Mostafazadeh-02-1, *Mostafazadeh-02-2, *Mostafazadeh-02-3}, which can lead to the real spectra of non-Hermitian systems (see Sec.~\ref{subsec: real spectra} for details). These symmetry is included in the 38-fold internal symmetry in non-Hermitian physics~\cite{KSUS-19, Bernard-LeClair-02}. Examples of the symmetry operators $\mathcal{T}$ and $\eta$ are given in the subsequent sections [see Eqs.~(\ref{eq: NH-BHZ-TRS}) and (\ref{eq: NH-BHZ-pH})].

The $\mathbb{Z}_{2}$ topological phase survives non-Hermiticity as long as reciprocity in Eq.~(\ref{eq: TRS}) is respected and the gap for the real part of eigenenergies remains open [i.e., $\forall\,{\bm k}~~\mathrm{Re}\,E \left( \bm{k}\right) \neq 0$; real line gap in Ref.~\cite{KSUS-19}]. Furthermore, even a $\mathbb{Z}$ topological invariant is well defined in the presence of additional pseudo-Hermiticity in Eq.~(\ref{eq: pH}). To see this $\mathbb{Z}$ invariant, let us focus on a matrix $\eta H \left( {\bm k} \right)$. Because of pseudo-Hermiticity in Eq.~(\ref{eq: pH}), $\eta H \left( {\bm k} \right)$ is Hermitian:
\begin{equation}
\left[ \eta H \left( {\bm k} \right) \right]^{\dag} = \eta H \left( {\bm k} \right).
\end{equation}
In addition, $\eta H \left( {\bm k} \right)$ has a gap when the original non-Hermitian Hamiltonian $H \left( \bm{k} \right)$ has a gap for the real part of eigenenergies. Consequently, the Chern number is well defined for $\eta H \left( \bm{k} \right)$, which characterizes the $\mathbb{Z}$ topological phase of $H \left( \bm{k} \right)$. This is contrasted with the vanishing Chern number for $H \left( \bm{k}\right)$ due to time-reversal symmetry (reciprocity). Notably, if reciprocity and pseudo-Hermiticity commute with each other (i.e., $\mathcal{T}\eta^{*} = \eta \mathcal{T}$) instead of Eq.~(\ref{eq: TRS - unitary}), $\eta H \left( \bm{k} \right)$ respects time-reversal symmetry and its Chern number vanishes. The $\mathbb{Z}$ topological phases protected by reciprocity in Eq.~(\ref{eq: TRS}) and pseudo-Hermiticity in Eq.~(\ref{eq: pH}) are consistent with the 38-fold classification of non-Hermitian topological phases (see Table~IX in Ref.~\cite{KSUS-19}, with the symmetry class ``AI$+ \eta_{-}$" and two dimensions). The combination of reciprocity and pseudo-Hermiticity enables the well-defined $\mathbb{Z}$ invariant, while we only have a $\mathbb{Z}_{2}$ invariant if pseudo-Hermiticity is not respected (see also Table~V in Ref.~\cite{KSUS-19}, with the symmetry class ``AII$^{\dag}$" and two dimensions). It is also remarkable that the $\mathbb{Z}$ invariant is equivalent to the time-reversal-invariant Chern number in Refs.~\cite{Sato-11, *Esaki-11, KSUS-19}.

\subsection{Real spectra}
	\label{subsec: real spectra}

A combination of the symmetry in Eqs.~(\ref{eq: TRS}) and (\ref{eq: pH}) leads to the entirely real spectra for both bulk and edges. The real spectra of the bulk are ensured by pseudo-Hermiticity in Eq.~(\ref{eq: pH}). To see this, let $E_{n} \left( \bm{k} \right)$ be an eigenenergy of $H \left( \bm{k} \right)$ and $\ket{u_{n} \left( \bm{k} \right)}$ ($| u_{n} \left( \bm{k} \right) \rangle\!\rangle$) be the corresponding right (left) eigenstate:
\begin{equation} \begin{split}
H \left( \bm{k} \right) \ket{u_{n} \left( \bm{k} \right)}
&= E_{n} \left( \bm{k} \right) \ket{u_{n} \left( \bm{k} \right)},\\
\langle\!\langle u_{n} \left( \bm{k} \right) | H \left( \bm{k} \right)
&= E_{n} \left( \bm{k} \right) \langle\!\langle u_{n} \left( \bm{k} \right) |.
\end{split} \end{equation}
In the presence of pseudo-Hermiticity in Eq.~(\ref{eq: pH}), we have
\begin{eqnarray}
H \left( \bm{k} \right) \left[ \eta | u_{n} \left( \bm{k} \right) \rangle\!\rangle \right]
&=& \eta H^{\dag} \left( \bm{k} \right) | u_{n} \left( \bm{k} \right) \rangle\!\rangle \nonumber \\
&=& E_{n}^{*} \left( \bm{k} \right) \left[ \eta | u_{n} \left( \bm{k} \right) \rangle\!\rangle \right],
\end{eqnarray}
which implies that $\eta | u_{n} \left( \bm{k} \right) \rangle\!\rangle$ is a right eigenstate of $H \left( \bm{k} \right)$ with the eigenenergy $E_{n}^{*} \left( \bm{k} \right)$. When non-Hermiticity is sufficiently weak, $\ket{u_{n} \left( \bm{k} \right)}$ and $\eta | u_{n} \left( \bm{k} \right) \rangle\!\rangle$ should coincide with each other since they are the same single state in the absence of non-Hermiticity. As a result, it holds
\begin{equation}
E_{n} \left( \bm{k} \right) = E_{n}^{*} \left( \bm{k} \right),\quad\mathrm{i.e.},\quad E_{n} \left( \bm{k} \right) \in \mathbb{R}.
\end{equation} 
On the other hand, when non-Hermiticity is strong enough to give rise to band touching, $\ket{u_{n} \left( \bm{k} \right)}$ and $\eta | u_{n} \left( \bm{k} \right) \rangle\!\rangle$ are different, so the corresponding eigenenergies become complex in a pair. Thus, even in the presence of non-Hermiticity, an energy band with a real spectrum remains real as long as it is isolated from other bands and pseudo-Hermiticity is preserved. It can have a complex spectrum only if the energy gap is closed.

On the other hand, pseudo-Hermiticity alone does not necessarily lead to the real spectra of the boundary states. This is because the boundary states are gapless and hence can have complex spectra. Nevertheless, their reality can be ensured by reciprocity in Eq.~(\ref{eq: TRS}). An important consequence of Eq.~(\ref{eq: TRS}) is Kramers degeneracy~\cite{Sato-11, *Esaki-11, KSUS-19}. To see this, we have
\begin{eqnarray}
H \left( \bm{k} \right) \left[ \mathcal{T} | u_{n}^{*} \left( - \bm{k} \right) \rangle\!\rangle \right]
&=& \mathcal{T} H^{T} \left( - \bm{k} \right) | u_{n}^{*} \left( - \bm{k} \right) \rangle\!\rangle \nonumber \\
&=& E_{n} \left( - \bm{k} \right) \left[ \mathcal{T} | u_{n}^{*} \left( - \bm{k} \right) \rangle\!\rangle \right],
\end{eqnarray}
which implies that $\mathcal{T} | u_{n}^{*} \left( - \bm{k} \right) \rangle\!\rangle$ is a right eigenstate of $H \left( \bm{k} \right)$ with the eigenenergy $E_{n} \left( - \bm{k} \right)$. Hence, at a time-reversal-invariant momentum ${\bm k}_{\rm TRIM}$ [i.e., $H \left( \bm{k}_{\rm TRIM} \right) = H \left( - \bm{k}_{\rm TRIM} \right)$], both $\ket{u_{n} \left( \bm{k}_{\rm TRIM} \right)}$ and $\mathcal{T} | u_{n}^{*} \left( \bm{k}_{\rm TRIM} \right) \rangle\!\rangle$ belong to the same eigenenergy $E_{n} \left( \bm{k}_{\rm TRIM} \right)$. Moreover, because of $\mathcal{T}^{T} = - \mathcal{T}$, we have
\begin{eqnarray}
&&\langle\!\langle u_{n} \left( \bm{k}_{\rm TRIM} \right) | \mathcal{T} | u_{n} \left( \bm{k}_{\rm TRIM} \right) \rangle\!\rangle \nonumber \\
&&\qquad= \langle\!\langle u_{n} \left( \bm{k}_{\rm TRIM} \right) | \mathcal{T}^{T} | u_{n} \left( \bm{k}_{\rm TRIM} \right) \rangle\!\rangle \nonumber \\
&&\qquad = - \langle\!\langle u_{n} \left( \bm{k}_{\rm TRIM} \right) | \mathcal{T} | u_{n} \left( \bm{k}_{\rm TRIM} \right) \rangle\!\rangle,
\end{eqnarray}
leading to 
\begin{equation}
\langle\!\langle u_{n} \left( \bm{k}_{\rm TRIM} \right) | \mathcal{T} | u_{n} \left( \bm{k}_{\rm TRIM} \right) \rangle\!\rangle = 0.
\end{equation} 
This indicates that $\ket{u_{n} \left( \bm{k}_{\rm TRIM} \right)}$ and $\mathcal{T} | u_{n}^{*} \left( \bm{k}_{\rm TRIM} \right) \rangle\!\rangle$ are biorthogonal~\cite{Brody-14} and linearly independent of each other. This Kramers degeneracy at time-reversal-invariant momenta is retained as long as reciprocity in Eq.~(\ref{eq: TRS}) is respected.

Now, suppose the Chern number of $\eta H \left( \bm{k} \right)$ is one. In the presence of Hermiticity, a pair of helical edge states appears and crosses at a time-reversal-invariant momentum. The bulk spectrum remains real because of pseudo-Hermiticity as long as the gap for the real part of the spectrum is open. On the other hand, the helical edge states are gapless and hence pseudo-Hermiticity alone cannot ensure their real spectrum. However, reciprocity and the consequent Kramers degeneracy ensure the real spectrum of the helical edge states. In fact, if the pair of the helical edge states mixed with each other and formed a complex-conjugate pair, Kramers degeneracy at the time-reversal-invariant momentum would be lifted, which is forbidden in the presence of reciprocity. Thus, the spectrum is entirely real for both bulk and edges as a consequence of the combination of pseudo-Hermiticity and reciprocity.

Next, suppose the Chern number of $\eta H \left( \bm{k} \right)$ is two. In contrast to the previous case, two pairs of helical edge states appear, and neither of them necessarily crosses at time-reversal-invariant momenta. No degeneracy is guaranteed away from time-reversal-invariant momenta even in the presence of reciprocity. As a result, the helical edge states can mix with each other and form complex-conjugate pairs with exceptional points. Still, the bulk spectrum is real as long as the gap for the real part of the spectrum remains open. Thus, the system supports two pairs of helical lasing edge states. A model of such a symmetry-protected topological laser is provided in Refs.~\cite{Sato-11, *Esaki-11, KSUS-19}.

Notably, the bulk spectrum can change according to boundary conditions. This is a unique feature of non-Hermitian systems called the non-Hermitian skin effect~\cite{Lee-16, Kunst-18, YW-18, *YSW-18}. However, when the bulk spectrum is real because of pseudo-Hermiticity (or parity-time symmetry), no skin effect occurs, i.e., the bulk spectrum under the periodic boundary conditions and that under the open boundary conditions always coincide with each other~\cite{Esaki-11, KSUS-19}.

\subsection{Complex spectra in $\mathbb{Z}_{2}$ topological insulators}
	\label{subsec: complex spectra}

Symmetry in Eqs.~(\ref{eq: TRS (H)}) and (\ref{eq: unitary (H)}) for Hermitian Hamiltonians can be respectively generalized to non-Hermitian systems in a different manner as
\begin{eqnarray}
\mathcal{T} H^{*} \left( \bm{k} \right) \mathcal{T}^{-1}
&=& H \left( - \bm{k} \right),\quad
\mathcal{T}\mathcal{T}^{*} = -1,
	\label{eq: TRS-cc} \\
\eta H \left( \bm{k} \right) \eta^{-1} 
&=& H \left( \bm{k} \right),\quad
\eta^{2} = 1.
	\label{eq: pH-nd}
\end{eqnarray}
In the presence of Hermiticity, Eqs.~(\ref{eq: TRS-cc}) and (\ref{eq: pH-nd}) respectively coincide with Eqs.~(\ref{eq: TRS}) and (\ref{eq: pH}), both of which reduce to Eqs.~(\ref{eq: TRS (H)}) and (\ref{eq: unitary (H)}). However, this is not the case for non-Hermitian Hamiltonians because of the distinction between complex conjugation and transposition [i.e., $H^{*} \left( \bm{k} \right) \neq H^{T} \left( \bm{k} \right)$]. Whereas time-reversal symmetry in Eq.~(\ref{eq: TRS-cc}) leads to Kramers degeneracy for eigenstates with real eigenenergies~\cite{KHGAU-19}, it results in no degeneracy for generic eigenstates with complex eigenenergies. This is contrasted with reciprocity in Eq.~(\ref{eq: TRS}), which ensures Kramers degeneracy for all the eigenstates with complex eigenenergies. Furthermore, symmetry in Eq.~(\ref{eq: pH-nd}) does not ensure the reality of the spectrum contrary to pseudo-Hermiticity in Eq.~(\ref{eq: pH}). Therefore, the other generalization in Eqs.~(\ref{eq: TRS-cc}) and (\ref{eq: pH-nd}) does not generally lead to the real spectra of non-Hermitian topological systems. 

For example, a non-Hermitian extension of the BHZ model with Eq.~(\ref{eq: TRS-cc}) is investigated in Appendix~\ref{appendix: BHZ-complex}. Because of the symmetry protection, the topological phase and the helical edge states survive even in the presence of non-Hermiticity. However, non-Hermiticity mixes these helical edge states and creates a pair of exceptional points, and the Kramers degeneracy at the time-reversal-invariant momentum is lifted. Consequently, the edge spectrum generally becomes complex. In contrast to this extension, another non-Hermitian extension of the BHZ model with Eq.~(\ref{eq: TRS}), which we consider in the subsequent sections, can possess entirely real spectra even in the presence of non-Hermiticity.

\section{Continuum Dirac Hamiltonian}
	\label{sec: Dirac}

Using non-Hermitian Dirac Hamiltonians with parity-time symmetry, Ref.~\cite{Hu-11} showed that the entirely real spectra of both bulk and edges are impossible. As discussed above, however, the discussion there is not directly applicable in the presence of additional symmetry such as pseudo-Hermiticity and reciprocity. To confirm this fact, we consider a non-Hermitian Dirac Hamiltonian and its spectrum in a similar manner to Ref.~\cite{Hu-11}. A non-Hermitian Dirac Hamiltonian having reciprocity in Eq.~(\ref{eq: TRS}) and pseudo-Hermiticity in Eq.~(\ref{eq: pH}) is generally described by
\begin{eqnarray}
&&H \left( {\bm k} \right)
= \left( \sigma_{z} + \ii \gamma \sigma_{x} \right) k_{x} \tau_{x} + k_{y} \tau_{y} + \Delta \tau_{z}.
	\label{eq: NH-Dirac}
\end{eqnarray}
Here, $\gamma \in \mathbb{R}$ describes the degree of non-Hermiticity, and $\Delta \in \mathbb{R}$ describes the mass parameter that determines the topological phases. This Dirac model indeed respects reciprocity in Eq.~(\ref{eq: TRS}) and pseudo-Hermiticity in Eq.~(\ref{eq: pH}) (i.e., $\mathcal{T} = \ii \sigma_{y}$ and $\eta = \sigma_{z}$):
\begin{eqnarray}
\left( \ii \sigma_{y} \right) H^{T} \left( \bm{k} \right) \left( \ii \sigma_{y} \right)^{-1}
&=& H \left( - \bm{k} \right),\quad
\left( \ii \sigma_{y} \right) \left( \ii \sigma_{y} \right)^{*} = -1,\qquad \label{eq: NH-BHZ-TRS}\\
\sigma_{z} H^{\dag} \left( \bm{k} \right) \sigma_{z}^{-1} 
&=& H \left( \bm{k} \right),\quad
\sigma_{z}^{2} = 1. \label{eq: NH-BHZ-pH}
\end{eqnarray}
The bulk spectrum is readily obtained as
\begin{equation}
E \left( \bm{k} \right) = \pm \sqrt{\left( 1-\gamma^{2} \right) k_{x}^{2} + k_{y}^{2} + \Delta^{2}},
\end{equation}
which is entirely real for $\left| \gamma \right| \leq 1$ as a direct consequence of pseudo-Hermiticity in Eq.~(\ref{eq: NH-BHZ-pH}). It is two-fold degenerate because of reciprocity in Eq.~(\ref{eq: NH-BHZ-TRS}).

Even though the bulk spectrum is entirely real, the edge spectrum is not necessarily real. In fact, Ref.~\cite{Hu-11} showed that non-Hermiticity mixes a pair of edge states and makes the edge spectrum complex in a large class of non-Hermitian topological insulators. Still, the Dirac Hamiltonian~(\ref{eq: NH-Dirac}) possesses the entirely real spectrum even for the edges because of additional pseudo-Hermiticity and reciprocity. To see this, we consider an interface across which topological phases change. We assume that the system is uniform along the $x$ direction and has a domain wall at $y = 0$. For the region $y > 0$ ($y < 0$), the mass parameter is assumed to be $\Delta \left( y \right) > 0$ [$\Delta \left( y \right) < 0$]. The corresponding continuum Hamiltonian reads
\begin{eqnarray}
H \left( k_{x}, y \right)
= \left( \sigma_{z} + \ii \gamma \sigma_{x} \right) k_{x} \tau_{x} - \ii \tau_{y} \frac{\partial}{\partial y} + \Delta \left( y \right) \tau_{z}.\quad
	\label{eq: contiuum-Hamiltonian-Delta-y}
\end{eqnarray}
For $k_{x} = 0$, a Kramers pair of zero-energy bound states appears around the interface $y=0$. Solving the Shr\"odinger equation
\begin{equation}
\left[ - \ii \tau_{y} \frac{\partial}{\partial y} + \Delta \left( y \right) \tau_{z} \right] \ket{\Psi_{\uparrow (\downarrow)}} = 0, 
\end{equation}
we have
\begin{equation}
\ket{\Psi_{\uparrow (\downarrow)}}
= e^{-\int_{0}^{y} \Delta\,( y' )\,dy'} \ket{\uparrow (\downarrow)} \ket{-},
\end{equation}
where $\ket{\uparrow (\downarrow)}$ and $\ket{-}$ are the eigenstates of $\sigma_{z}$ and $\tau_{x}$, respectively [i.e., $\sigma_{z} \ket{\uparrow (\downarrow)} = + \left( - \right) \ket{\uparrow (\downarrow)}$ and $\tau_{x} \ket{-} = - \ket{-}$]. Away from the time-reversal-invariant momentum $k_{x} = 0$, these boundary states have nonzero eigenenergies, which form the energy dispersion of the helical boundary states. The effective boundary Hamiltonian around $k_{x} = 0$ is obtained as
\begin{eqnarray}
H_{\rm edge} \left( k_{x}, y \right) 
&\simeq& \left( \begin{array}{@{\,}cc@{\,}} 
	\braket{\Psi_{\uparrow} | H | \Psi_{\uparrow}} & \braket{\Psi_{\downarrow} | H | \Psi_{\uparrow}} \\
	\braket{\Psi_{\uparrow} | H | \Psi_{\downarrow}} & \braket{\Psi_{\downarrow} | H | \Psi_{\downarrow}} \\ 
	\end{array} \right) \nonumber \\
&=& e^{-2\int_{0}^{y} \Delta\,( y' )\,dy'} \left( \sigma_{z} + \ii \gamma \sigma_{x} \right) k_{x}.
\end{eqnarray}
The energy dispersion is given as
\begin{equation}
E_{\rm edge} \left( k_{x} \right) = \pm \sqrt{1 - \gamma^{2}}\,k_{x},
	\label{eq: edge-spectrum-continuum}
\end{equation}
which is indeed real for $\left| \gamma \right| \leq 1$.

We again stress that Kramers degeneracy plays a crucial role in the reality of the boundary spectrum. In the absence of reciprocity in Eq.~(\ref{eq: NH-BHZ-TRS}), the Kramers degeneracy at $k_{x} = 0$ is lifted by non-Hermitian perturbations and the boundary spectrum becomes complex, as discussed in Ref.~\cite{Hu-11}. In the presence of reciprocity, by contrast, the Kramers degeneracy cannot be lifted and the boundary spectrum remains real.

\section{Non-Hermitian Bernevig-Hughes-Zhang model}
	\label{sec: BHZ}

\subsection{Model and symmetry}

As a prime example of the preceding discussions, we consider a non-Hermitian extension of the BHZ model. The Hamiltonian in momentum space is given as
\begin{eqnarray}
&&H_{\rm BHZ} \left( {\bm k} \right)
= \left( m + t \cos k_{x} + t \cos k_{y} \right) \tau_{z} + t \left( \sin k_{y} \right) \tau_{y} \nonumber \\
&&\qquad\qquad\qquad\quad + t \left( \sin k_{x} \right) \sigma_{z} \tau_{x} + \ii \gamma \left( \sin k_{x} \right) \sigma_{x} \tau_{x},
	\label{eq: NH-BHZ}
\end{eqnarray}
where $t, m, \gamma \in \mathbb{R}$ are the hopping amplitude, the mass parameter, and the degree of non-Hermiticity, respectively. We assume $t, \gamma \geq 0$ without loss of generality. In the absence of non-Hermiticity (i.e., $\gamma = 0$), Eq.~(\ref{eq: NH-BHZ}) reduces to the original Hermitian BHZ model in Eq.~(\ref{eq: H-BHZ}).

Around the time-reversal-invariant momentum $\bm{k} = 0$, the non-Hermitian BHZ model $H_{\rm BHZ} \left( \bm{k} \right)$ reduces to the continuum Dirac model in Sec.~\ref{sec: Dirac} (i.e., $t=1$ and $\Delta = m+2t$). It respects reciprocity in Eq.~(\ref{eq: NH-BHZ-TRS}) and pseudo-Hermiticity in Eq.~(\ref{eq: NH-BHZ-pH}). In addition, it respects parity (spatial-inversion) symmetry:
\begin{equation}
\tau_{z} H \left( \bm{k} \right) \tau_{z}^{-1}
= H \left( - \bm{k} \right),\quad
\tau_{z}^{2} = 1.
	\label{eq: parity}
\end{equation}
As a combination of these symmetry, $H_{\rm BHZ} \left( \bm{k} \right)$ also respects parity-time symmetry:
\begin{equation}
\left( \tau_{z} \sigma_{x} \right) H^{*} \left( \bm{k} \right) \left( \tau_{z} \sigma_{x} \right)^{-1}
= H \left( \bm{k} \right),\quad
\left( \tau_{z} \sigma_{x} \right)^{2} = 1.
	\label{eq: PT}
\end{equation}
While reciprocity and pseudo-Hermiticity are internal symmetry, parity symmetry and parity-time symmetry are spatial symmetry, the latter of which is fragile against disorder.

\subsection{Phase diagram}

The spectrum of $H_{\rm BHZ} \left( \bm{k} \right)$ is obtained as 
\begin{eqnarray}
&& E \left( {\bm k} \right)
= \pm \left[ \left( m + t \cos k_{x} + t \cos k_{y} \right)^{2} \right. \nonumber \\
&&\qquad\qquad\qquad+ \left( t^{2} - \gamma^{2} \right) \sin^{2} k_{x} + t^{2} \sin^{2} k_{y} \Bigr]^{1/2}.\quad
	\label{eq: BHZ spectrum}
\end{eqnarray}
A topological phase persists as long as a gap for the real part of eigenenergies is open [i.e., $\forall\,{\bm k}~~\mathrm{Re}\,E \left( \bm{k} \right) \neq 0$]; vanishing the real part of eigenenergies [i.e., $\exists\,{\bm k}~~\mathrm{Re}\,E \left( \bm{k} \right) = 0$] can be considered to be a topological phase transition. Here, $E \left( \bm{k} \right)$ in Eq.~(\ref{eq: BHZ spectrum}) is either real or purely imaginary. In particular, $E \left( \bm{k} \right)$ is always real for the time-reversal-invariant momenta $\bm{k}_{\rm TRIM} \in \{ \left( 0, 0\right), \left( 0, \pi\right), \left( \pi, 0\right), \left( \pi, \pi\right) \}$. Thus, if an energy gap for the real part of the spectrum is closed, it holds $E \left( \bm{k} \right) = 0$ for some $\bm{k}$, and vice versa. This reduces to the following gapless conditions according to $t$ and $\gamma$.

\begin{figure}[t]
\centering
	\includegraphics[width = 86mm]{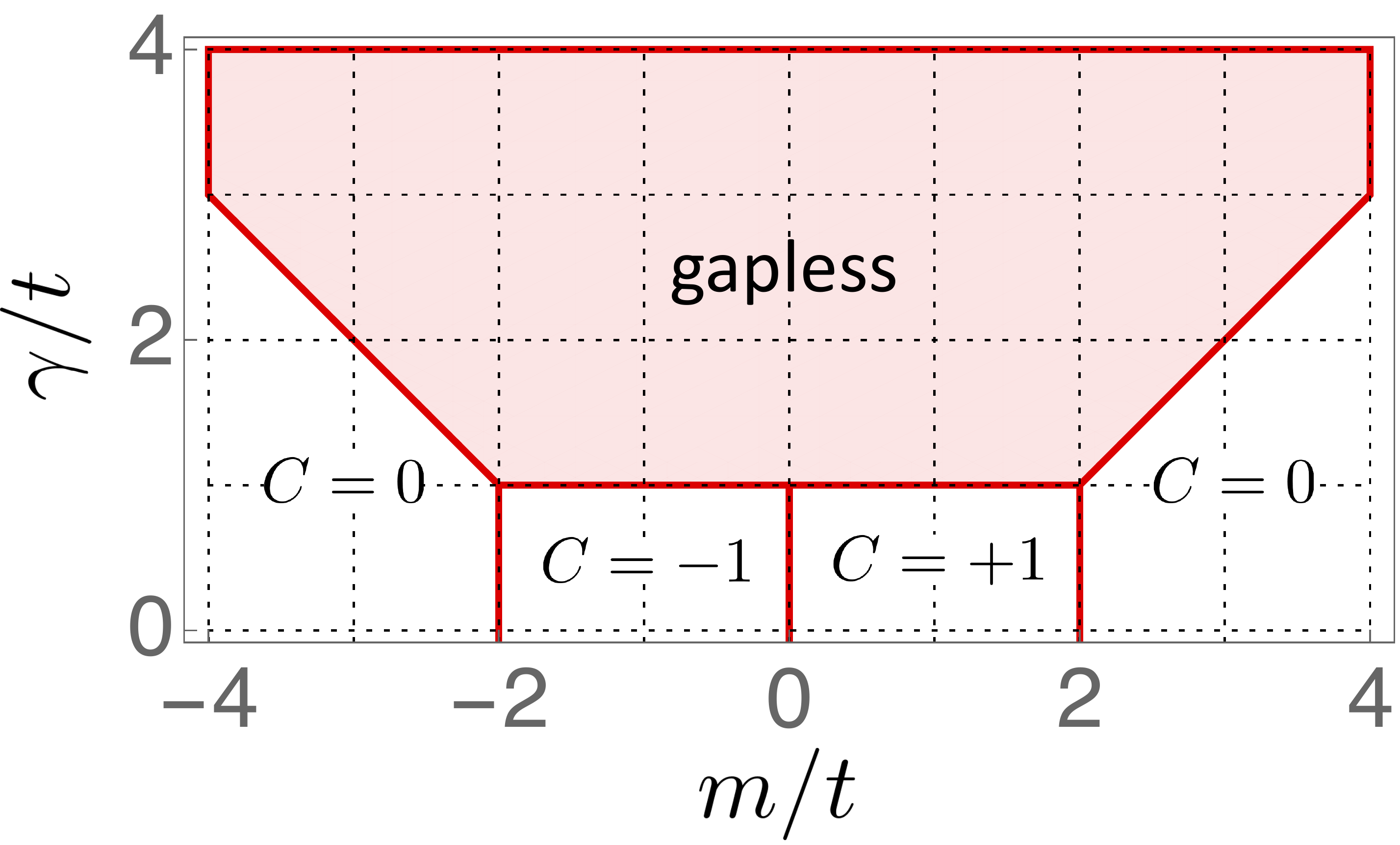} 
	\caption{Phase diagram of the non-Hermitian Bernevig-Hughes-Zhang model. Topological phase transitions occur at the phase boundaries, at which an energy gap for the real part of the complex spectrum closes. Each gapped phase is characterized by the Chern number $C \in \mathbb{Z}$ of $\eta H \left( \bm{k} \right)$. A pair of helical edge states appears for $\left| C \right| = 1$, whereas no edge states appear for $C = 0$.}
	\label{fig: phase diagram}
\end{figure}

\begin{itemize}
\item[(1)] $\gamma < t$.\,--- Since we have
\begin{equation} \begin{split}
&\left( m + t \cos k_{x} + t \cos k_{y} \right)^{2} \geq 0, \\
&\left( t^{2} - \gamma^{2} \right) \sin^{2} k_{x} \geq 0,\quad
t^{2} \sin^{2} k_{y} \geq 0,
\end{split} \end{equation}
$E \left( \bm{k} \right) = 0$ leads to
\begin{equation}
m + t \cos k_{x} + t \cos k_{y} = \sin k_{x} = \sin k_{y} = 0.
\end{equation}
Hence, we have
\begin{equation}
\begin{cases}
m = -2t & \mathrm{for}\quad\bm{k}_{0} = \left( 0, 0\right), \\
m = 0 & \mathrm{for}\quad\bm{k}_{0} = \left( 0, \pi\right), \left( \pi, 0 \right), \\
m = 2t & \mathrm{for}\quad\bm{k}_{0} = \left( \pi, \pi \right), \\
\end{cases}
\end{equation}
where $\bm{k}_{0}$ is a momentum satisfying $E \left( \bm{k}_{0} \right) = 0$.

\item[(2)] $\gamma = t$.\,--- Since $E \left( \bm{k} \right) = 0$ leads to
\begin{equation}
m + t \cos k_{x} + t \cos k_{y} = \sin k_{y} = 0,
\end{equation}
we have
\begin{equation}
\begin{cases}
-2t \leq m \leq 0 & \mathrm{for}\quad\bm{k}_{0} = \left( \mathrm{arccos} \left( 1+m/t\right), 0\right), \\
0 \leq m \leq 2t & \mathrm{for}\quad\bm{k}_{0} = \left( \mathrm{arccos} \left( 1-m/t\right), \pi\right). \\
\end{cases}
\end{equation}

\item[(3)] $\gamma > t$.\,--- Since we have
\begin{equation} \begin{split}
&E^{2} \left( 0, 0 \right) = \left( m+2t \right)^{2} \geq 0, \\
&E^{2} \left( 0, \pi \right) = E^{2} \left( \pi, 0 \right) = m^{2} \geq 0, \\
&E^{2} \left( \pi, \pi \right) = \left( m-2t \right)^{2} \geq 0, \\
\end{split} \end{equation}
there exists $\bm{k}_{0}$ satisfying $E \left( \bm{k}_{0} \right) = 0$ if and only if the minimum of $E^{2} \left( \bm{k} \right)$ is nonpositive. Then, we have
\begin{eqnarray}
&&E^{2} \left( \bm{k} \right)
= 2t \left( m+t\cos k_{x} \right) \cos k_{y} \nonumber \\
&&\qquad + \left( m+t\cos k_{x}\right)^{2} + \left( t^{2} - \gamma^{2} \right) \sin^{2} k_{x} + t^{2},
\end{eqnarray}
which implies that $E^{2} \left( \bm{k} \right)$ is minimum for $k_{y} = 0$ or $k_{y} = \pi$. Now, $E^{2} \left( k_{x}, 0 \right)$ is given as
\begin{eqnarray}
&&E^{2} \left( k_{x}, 0 \right)
= \gamma^{2} \left[ \cos k_{x} + \frac{t \left( m+t \right)}{\gamma^{2}} \right]^{2} \nonumber \\
&&\qquad\qquad\qquad\quad + \left( 1 - \frac{t^{2}}{\gamma^{2}} \right) \left[ \left( m+t \right)^{2} - \gamma^{2} \right],
\end{eqnarray}
and $E^{2} \left( k_{x}, 0 \right)$ is nonnegative for $k_{x} = 0$ and $k_{x} = \pi$. Thus, we have $E \left( \bm{k}_{0} \right) = 0$ for 
\begin{equation}
\bm{k}_{0} = \left( \mathrm{arccos} \left( -\frac{t \left( m+t\right)}{\gamma^{2}} \right), 0 \right)
\end{equation}
if and only if 
\begin{equation}
\left| \frac{t \left( m+t \right)}{\gamma^{2}} \right| \leq 1,\quad \left( m+t \right)^{2} - \gamma^{2} < 0
\end{equation}
are satisfied; these inequalities reduce to $\gamma > \left| m+t \right|$. Similarly, we have $E \left( \bm{k}_{0} \right) = 0$ for \begin{equation}
\bm{k}_{0} = \left( \mathrm{arccos} \left( - \frac{t \left( m-t \right)}{\gamma^{2}} \right), \pi \right)
\end{equation}
as long as $\gamma > \left| m-t \right|$ is satisfied.

\end{itemize}

\begin{figure*}[t]
\centering
\includegraphics[width=172mm]{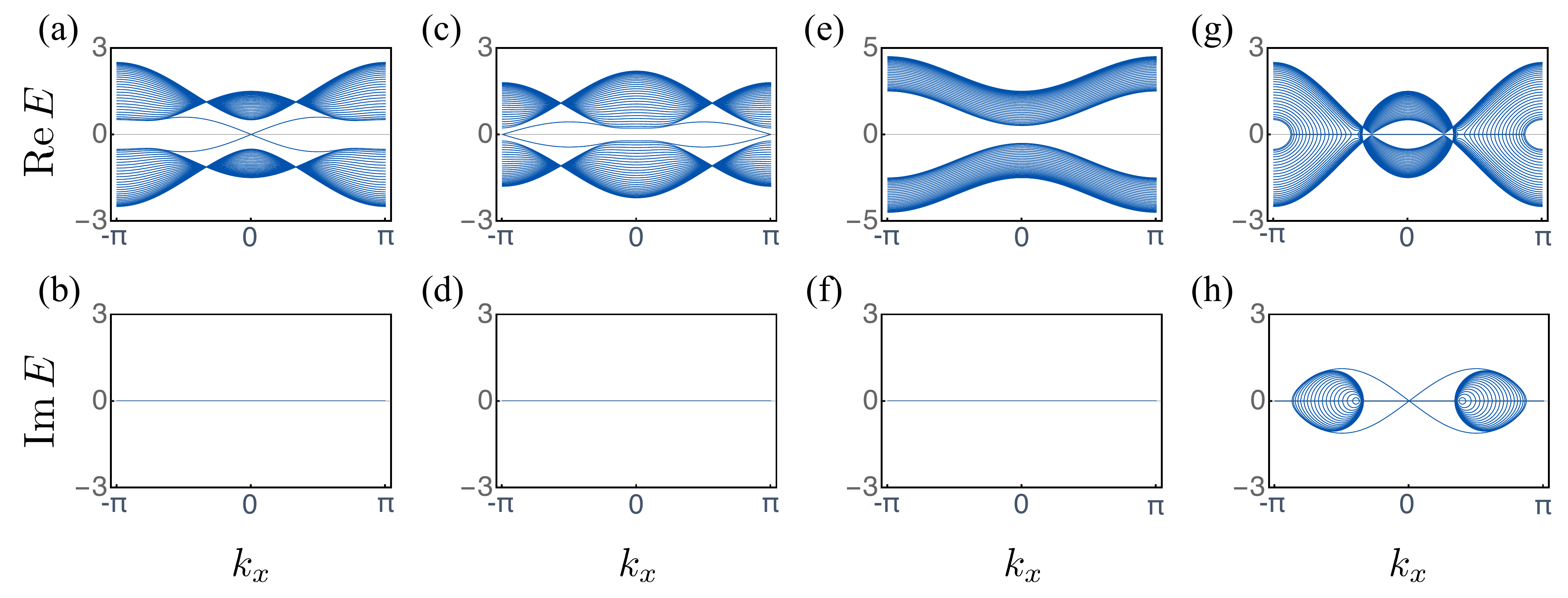} 
\caption{Complex spectrum of the non-Hermitian Bernevig-Hughes-Zhang model. The open boundary conditions are imposed in the $y$ direction ($30$ sites), whereas the periodic boundary conditions are imposed in the $x$ direction, along which the wavenumber $k_{x}$ is defined. (a, b)~Gapped and topologically nontrivial phase ($t = 1.0$, $m = -0.5$, $\gamma = 0.8$; $C = -1$). A pair of helical edge states appears around $k_{x} = 0$. (c, d)~Gapped and topologically nontrivial phase ($t = 1.0$, $m = 0.2$, $\gamma = 0.9$; $C = +1$). A pair of helical edge states appears around $k_{x} = \pm \pi$. (e, f)~Gapped and topologically trivial phase ($t = 1.0$, $m = -2.5$, $\gamma = 1.0$; $C = 0$). No edge states appear between the gapped bands. (g, h)~Gapless phase ($t = 1.0$, $m = -0.5$, $\gamma = 1.5$). The spectrum is entirely real in the gapped phases (a-f), but it is complex in the gapless phase (g, h).}
	\label{fig: spectrum}
\end{figure*}

The obtained phase diagram is provided in Fig.~\ref{fig: phase diagram}. Since topology is invariant unless an energy gap is closed, the topological invariant in each gapped phase is obtained by continuously deforming the non-Hermitian system into the corresponding Hermitian system without closing the energy gap. In the absence of non-Hermiticity (i.e., $\gamma = 0$), we have
\begin{eqnarray}
&&\eta H_{\rm BHZ} \left( {\bm k} \right)
= \left( m + t \cos k_{x} + t \cos k_{y} \right) \tau_{z} \sigma_{z} \nonumber \\
&&\qquad\qquad\qquad\quad + t \left( \sin k_{y} \right) \tau_{y} \sigma_{z} + t \left( \sin k_{x} \right) \tau_{x}.\quad
\end{eqnarray}
The Chern number $C$ of $\eta H_{\rm BHZ} \left( \bm{k} \right)$ with $\gamma = 0$ is readily obtained as
\begin{equation}
C = \begin{cases}
\mathrm{sgn} \left( m\right) &\mathrm{for}~\left| m/t \right| < 2,\\
~~~~0 & \mathrm{for}~\left| m/t \right| > 2.\\
\end{cases}
\end{equation}
This Chern number $C$ is the topological invariant of $H_{\rm BHZ} \left( \bm{k} \right)$ in the gapped phases, as shown in Fig.~\ref{fig: phase diagram}.

\subsection{Helical edge states}
	\label{sec: helical edge states}

Corresponding to the nontrivial topology of the bulk, a pair of helical edge states appears under the open boundary conditions. We here investigate the non-Hermitian BHZ model with periodic boundaries in the $x$ direction and open boundaries in the $y$ direction:
\begin{eqnarray}
&&\hat{H}_{\rm BHZ} = \sum_{k_{x}, y} \left\{ \left[ \hat{c}_{k_{x}, y+1}^{\dag} \frac{t \left( \tau_{z} + \ii \tau_{y} \right)}{2} \hat{c}_{k_{x}, y} + {\rm H.c.} \right] \right. \nonumber \\
&&\qquad\qquad +~\hat{c}_{k_{x}, y}^{\dag} \left[ \left( m + t \cos k_{x} \right) \tau_{z} + t \left( \sin k_{x} \right) \sigma_{z} \tau_{x} 
\right. \nonumber \\
&& \left.\qquad\qquad\qquad\qquad\qquad+\ii \gamma \left( \sin k_{x}\right) \sigma_{x} \tau_{x} \right] \hat{c}_{k_{x}, y} \biggr\},
\end{eqnarray}
where $\hat{c}_{k_{x}, y}$ ($\hat{c}_{k_{x}, y}^{\dag}$) annihilates (creates) a particle with momentum $k_{x}$ and on site $y$ that has four internal degrees of freedom. The spectrum is shown in Fig.~\ref{fig: spectrum}. In the gapped phases with nontrivial topology, a pair of helical edge states indeed appears at both edges [Fig.~\ref{fig: spectrum}\,(a-d)]. On the other hand, no edge states appear in the gapped phase with trivial topology [Fig.~\ref{fig: spectrum}\,(e,~f)]. The spectra are entirely real even in the presence of the edge states. When non-Hermiticity is sufficiently strong and the gap for the real part of the spectrum closes, the bulk spontaneously breaks pseudo-Hermiticity and its spectrum becomes complex [Fig.~\ref{fig: spectrum}\,(g,~h)]. 

We note that no skin effects occur in $H_{\rm BHZ} \left( \bm{k} \right)$. Thus, similar results are obtained under different types of the open boundary conditions, i.e., the open boundary conditions in the $x$ direction and the periodic boundary conditions in the $y$ direction, or the open boundary conditions in both $x$ and $y$ directions. This is contrasted with non-Hermitian systems that exhibit skin effects, including non-Hermitian Chern insulators~\cite{YSW-18, KSU-18}.

The energy dispersions and wavefunctions of the helical edge states are analytically obtained in the following manner. Let us consider a pair of helical edge states localized around $y = 1$. The edge states are denoted as
\begin{equation}
\hat{\Psi}_{\rm edge} \propto \sum_{y} \lambda^{y-1}~( \hat{c}_{k_{x}, y}^{\dag} \vec{v} ),
\end{equation}
where $\lambda$ is a parameter that determines the localization length [given by $- \left( \log \left| \lambda \right|\right)^{-1}$], and $\vec{v}$ is a four-component vector that describes the internal degrees of freedom. Then, the Schr\"odinger equation $[ \hat{H}, \hat{\Psi}_{\rm edge} ] = E_{\rm edge} \hat{\Psi}_{\rm edge}$ reduces to
\begin{equation}
\left( \lambda^{-1} T + M + \lambda T^{\dag} \right) \vec{v} = E_{\rm edge}\,\vec{v}
	\label{eq: Sch bulk}
\end{equation}
in the bulk and 
\begin{equation}
\left( M + \lambda T^{\dag} \right) \vec{v} = E_{\rm edge}\,\vec{v}
	\label{eq: Sch edge}
\end{equation}
at the edge. Here, $T$ and $M$ are defined as
\begin{equation} \begin{split}
T &:= \frac{t \left( \tau_{z} + \ii \tau_{y} \right)}{2},\\
M &:= \left( m + t \cos k_{x} \right) \tau_{z} + t \left( \sin k_{x} \right) \tau_{x} \sigma_{z} + \ii \gamma \left( \sin k_{x}\right) \tau_{x} \sigma_{x}.
\end{split} \end{equation}
In addition, we take the semi-infinite limit and neglect the effect of the other edge. Equations~(\ref{eq: Sch bulk}) and (\ref{eq: Sch edge}) lead to $T\,\vec{v} = 0$, which implies
\begin{equation}
\vec{v} = \begin{pmatrix}
\vec{v}_{\sigma} \\ -\vec{v}_{\sigma}
\end{pmatrix},
\end{equation}
with a two-component vector $\vec{v}_{\sigma}$ that acts in the space of $\sigma_{i}$'s. Using Eq.~(\ref{eq: Sch bulk}) or Eq.~(\ref{eq: Sch edge}), we have
\begin{eqnarray}
\left( \lambda t + m + t \cos k_{x}\right) \vec{v}_{\sigma} &=& 0, \label{eq: lambda} \\
\left[ t \left( \sin k_{x} \right) \sigma_{z} + \ii \gamma \left( \sin k_{x} \right) \sigma_{x} \right] \vec{v}_{\sigma} &=& -E_{\rm edge} \vec{v}_{\sigma}. \label{eq: E-edge}
\end{eqnarray}
Since $\vec{v}_{\sigma}$ is nonvanishing, Eq.~(\ref{eq: lambda}) leads to
\begin{equation}
\lambda = - \frac{m}{t} - \cos k_{x},
\end{equation}
which determines the localization length of the helical edge states. Here, $\lambda$ should be less than $1$ so that the edge states can be normalized. This gives $\left| m/t + \cos k_{x}\right| < 1$. For the presence of the helical edge states, there exists a wavenumber $k_{x}$ that satisfies this inequality, which then leads to $\left| m/t \right| < 2$. This condition is compatible with the phase diagram in Fig.~\ref{fig: phase diagram}. Furthermore, Eq.~(\ref{eq: E-edge}) implies that $\vec{v}_{\sigma}$ is an eigenstate of the $2 \times 2$ matrix $t \left( \sin k_{x} \right) \sigma_{z} + \ii \gamma \left( \sin k_{x} \right) \sigma_{x}$ with the eigenenergy $-E_{\rm edge}$, which gives
\begin{equation}
E_{\rm edge} \left( k_{x} \right) = \pm \sqrt{t^{2} -\gamma^{2}} \sin k_{x}.
\end{equation}
Thus, the spectrum of the helical edge states is indeed real for $\gamma < t$. The obtained analytical results are consistent with the numerical results in Fig.~\ref{fig: spectrum}, as well as the results for the continuum Dirac Hamiltonian in Sec.~\ref{sec: Dirac}.

\subsection{Robustness to disorder}
	\label{sec: disorder}

\begin{figure}[t]
\centering
\includegraphics[width=86mm]{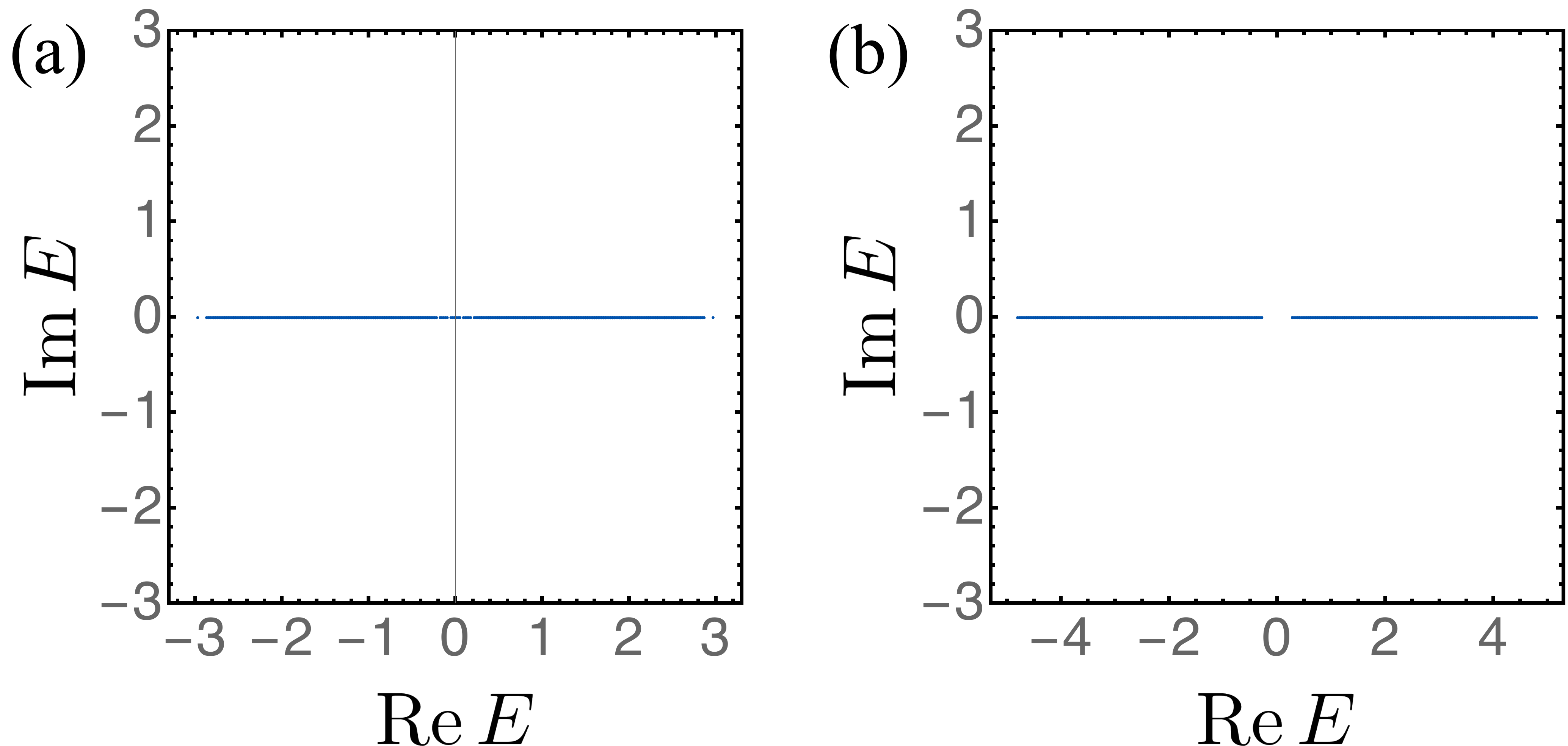} 
\caption{Complex spectrum of the non-Hermitian Bernevig-Hughes-Zhang model with disorder. The open boundary conditions are imposed in both $x$ and $y$ directions ($30 \times 30$ sites). Even in the presence of disorder, the spectrum is entirely real for both (a)~topological phase ($t = 1.0$, $m_{x, y} = -0.5 + 2.0\,\epsilon_{x, y}$, $\gamma = 0.8$) and (b)~trivial phase ($t = 1.0$, $m_{x, y} = -2.5 + 2.0\,\epsilon_{x, y}$, $\gamma = 1.0$). Here, $\epsilon_{x, y}$ is a random variable uniformly distributed over $\left[ -0.5, 0.5 \right]$.}
	\label{fig: disorder}
\end{figure}

The entirely real spectra in the non-Hermitian BHZ model are robust to disorder. To see this, we investigate the following disordered model:
\begin{eqnarray}
&&\hat{H}_{\rm BHZ} = \sum_{x, y} \left\{ \left[ \hat{c}_{x, y+1}^{\dag} \frac{t \left( \tau_{z} + \ii \tau_{y} \right)}{2} \hat{c}_{x, y} + {\rm H.c.} \right] \right. \nonumber \\
&&\qquad\qquad+ \left[ \hat{c}_{x+1, y}^{\dag} \frac{t \left( \tau_{z} + \ii \sigma_{z} \tau_{x} \right) - \gamma \sigma_{x} \tau_{x}}{2} \hat{c}_{x, y} \right. \nonumber \\
&&\qquad\qquad\qquad \left. + \hat{c}_{x, y}^{\dag} \frac{t \left( \tau_{z} - \ii \sigma_{z} \tau_{x} \right) + \gamma \sigma_{x} \tau_{x}}{2} \hat{c}_{x+1, y} \right] \nonumber \\
&&\qquad\qquad\qquad\qquad\qquad\quad+ \left. \hat{c}_{x, y}^{\dag} \left( m_{x, y} \tau_{z} \right) \hat{c}_{x, y} \right\},
	\label{eq: Hamiltonian-disorder}
\end{eqnarray}
where the open boundary conditions are imposed in both $x$ and $y$ directions. In contrast to the clean model, the mass parameters $m_{x, y}$ depend on the lattice sites $x, y$. As shown in Fig.~\ref{fig: disorder}, the spectrum of this disordered model is entirely real even in the presence of disorder. There, $m_{x, y}$'s are uniformly-distributed random variables. Such disorder breaks parity symmetry in Eq.~(\ref{eq: parity}) and parity-time symmetry in Eq.~(\ref{eq: PT}). On the other hand, reciprocity and pseudo-Hermiticity remain to be respected since they are internal symmetry.

In a similar manner to the clean model discussed in Sec.~\ref{subsec: real spectra}, the reality of the bulk spectrum is due to pseudo-Hermiticity. However, the discussion in Sec.~\ref{subsec: real spectra} is not directly applicable to the reality of the edge spectrum since it relies on translation invariance. Still, the real edge spectrum can be partially understood on the basis of the continuum models in Sec.~\ref{sec: Dirac}. Suppose the system includes disorder solely along the $y$ direction, and translation invariance is respected along the $x$ direction. Then, the space-dependent mass parameter $\Delta \left( y \right)$ of the continuum model in Eq.~(\ref{eq: contiuum-Hamiltonian-Delta-y}), which corresponds to $m_{x, y}$ of the lattice model in Eq.~(\ref{eq: Hamiltonian-disorder}), only changes the eigenstates and has no effect on the spectrum, as shown in Eq.~(\ref{eq: edge-spectrum-continuum}).

It is also notable that disorder generally tends to give rise to real spectra and stabilize non-Hermitian systems. Prime examples include the Hatano-Nelson model~\cite{Hatano-Nelson-96, *Hatano-Nelson-97}. It is a time-reversal-invariant system in one dimension whose hopping amplitudes exhibit asymmetry as the degree of non-Hermiticity. Because of this non-Hermiticity, it possesses the complex spectrum in the absence of disorder. In the presence of disorder, by contrast, some eigenstates are localized and uncorrelated with other eigenstates. Consequently, these localized eigenstates have real eigenenergies. This disorder-induced real spectrum is stable against many-body interaction~\cite{Hamazaki-19}. Thus, it is intuitively expected that disorder leads to the real spectra also in the non-Hermitian BHZ model, although symmetry or topology may change this behavior even qualitatively.

\section{Power oscillation}
	\label{sec: power oscillation}
	
Even when a non-Hermitian system possesses an entirely real spectrum, it exhibits unique phenomena that have no analogs in Hermitian systems. Eigenstates of a non-Hermitian Hamiltonian are biorthogonal to each other~\cite{Brody-14}:
\begin{equation}
\langle\!\langle u_{m} | u_{n} \rangle
\propto \delta_{mn},\quad
\langle  u_{m} | u_{n} \rangle\!\rangle
\propto \delta_{mn},
\end{equation}
where $\ket{u_{n}}$ ($| u_{n} \rangle\!\rangle$) is a right (left) eigenstate of the non-Hermitian Hamiltonian $H$. Nevertheless, they are in general nonorthogonal to each other:
\begin{equation}
\braket{u_{m} | u_{n}} \neq \delta_{mn},\quad
\langle\!\langle u_{m} | u_{n} \rangle\!\rangle \neq \delta_{mn}.
\end{equation}
An immediate physical consequence of the nonorthogonality between eigenstates is power oscillation. This is the oscillation of the norm (power) unique to non-Hermitian systems. When a wavefunction is initially prepared to be 
\begin{equation}
\ket{\psi \left( 0 \right)}
= \sum_{n} c_{n} \ket{u_{n}},\quad
c_{n} := \frac{\langle\!\langle u_{n} | \psi \left( 0 \right) \rangle}{\langle\!\langle u_{n} | u_{n} \rangle},
\end{equation}
it evolves into
\begin{equation}
\ket{\psi \left( t \right)}
= e^{-\ii Ht} \ket{\psi \left( 0 \right)}
= \sum_{n} c_{n} e^{-\ii E_{n} t}\ket{u_{n}},
\end{equation}
where $E_{n}$ is the eigenenergy that corresponds to $\ket{u_{n}}$ and $| u_{n} \rangle\!\rangle$. Its norm is given by 
\begin{equation}
\braket{\psi \left( t \right) | \psi \left( t \right)}
= \sum_{m, n} c_{m}^{*} c_{n} e^{\ii \left( E_{m}^{*} - E_{n} \right) t} \braket{u_{m} | u_{n}}.
\end{equation}
In Hermitian systems, this reduces to 
\begin{equation}
\braket{\psi \left( t \right) | \psi \left( t \right)} = \sum_{n} \left| c_{n} \right|^{2} \braket{u_{n} | u_{n}} = \braket{\psi \left( 0 \right) | \psi \left( 0 \right)}
\end{equation} 
because of the orthogonality between eigenstates (i.e., $\braket{u_{m} | u_{n}} \propto \delta_{mn}$) and the reality of eigenenergies (i.e., $E_{n}^{*} = E_{n}$). In non-Hermitian systems, by contrast, eigenstates are in general nonorthogonal, and hence the norm $\braket{\psi \left( t \right) | \psi \left( t \right)}$ depends on time, which is a clear manifestation of nonunitarity of the dynamics resulting from coupling to an external environment. Notably, even when eigenenergies are entirely real, eigenstates are still nonorthogonal and the norm oscillates in contrast to unitary dynamics of Hermitian systems. This power oscillation was experimentally observed in the bulk of an open photonic lattice with balanced gain and loss~\cite{Regensburger-12}. A quantum counterpart arises as oscillation of quantum information flow between a system and its environment~\cite{KAU-17}, which was observed in dissipative single photons~\cite{Xiao-19}. Furthermore, we note in passing that the power oscillation has an analogy with the norm leakage in open chaotic systems~\cite{Savin-97}.

\begin{figure}[t]
\centering
	\includegraphics[width = 72mm]{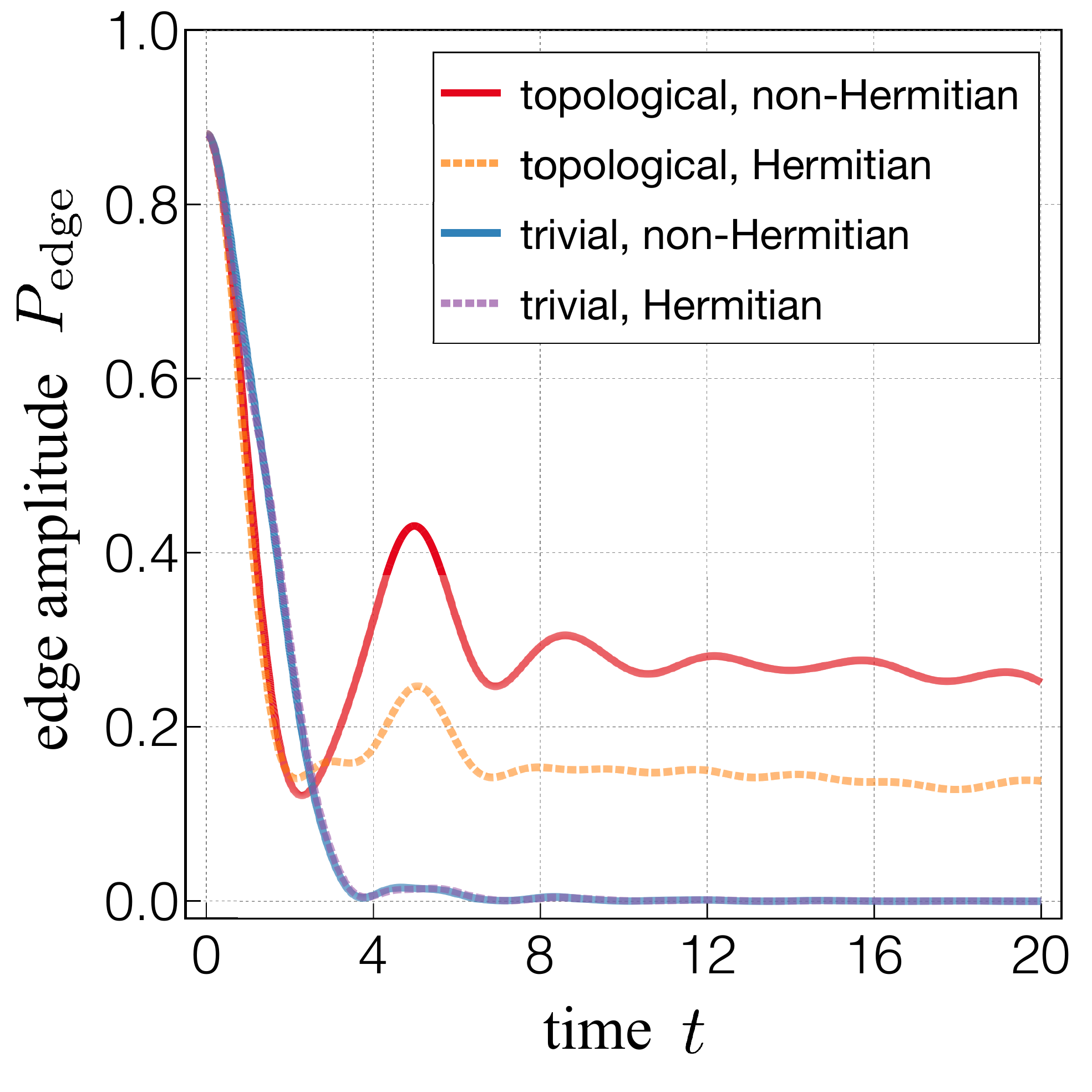} 
	\caption{Power oscillation at an edge in the non-Hermitian Bernevig-Hughes-Zhang model. An initial state is prepared to be a localized wavefunction $\ket{\psi \left( 0 \right)} \propto \sum_{x, y} e^{-\left( x-1\right)^{2}/36 - \left( y-1 \right)^{2}} \ket{x} \ket{y}$, and the evolutions of the amplitude at the edge [i.e., $P_{\rm edge} \left( t \right) := \left| \braket{y = 1 | e^{-\ii H_{\rm BHZ} t} | \psi \left( 0 \right)} \right|^{2}$] are shown. The two-dimensional system consists of $30 \times 30$ sites, and has periodic boundaries in the $x$ direction and open boundaries in the $y$ direction. The red solid curve shows the dynamics for the non-Hermitian topological phase ($t = 1.0$, $m = -0.5$, $\gamma = 0.8$), whereas the blue solid curve shows the dynamics for the non-Hermitian trivial phase ($t = 1.0$, $m = -2.5$, $\gamma = 1.0$); the orange dotted curve shows the dynamics for the Hermitian topological phase ($t = 1.0$, $m = -0.5$, $\gamma = 0$), whereas the violet dotted curve shows the dynamics for the Hermitian trivial phase ($t = 1.0$, $m = -2.5$, $\gamma = 0$).}
	\label{fig: power oscillation}
\end{figure}

The helical edge states oscillate in the non-Hermitian BHZ model. As an illustration, we investigate the non-Hermitian BHZ model $H_{\rm BHZ}$ with periodic boundaries in the $x$ direction and open boundaries in the $y$ direction, in a similar manner to Sec.~\ref{sec: helical edge states}. The number of sites is $L_{x} \times L_{y}$. An eigenenergy and the corresponding right (left) eigenstate of $H_{\rm BHZ} \left( k_{x} \right)$ are respectively denoted as $E_{n} \left( k_{x} \right)$ and $\ket{u_{n} \left( k_{x} \right)}$ ($| u_{n} \left( k_{x} \right) \rangle\!\rangle$) with $n=1,2,\cdots, 4L_{y}$, where $H_{\rm BHZ} \left( k_{x} \right)$ is a Fourier transform of the original Hamiltonian $H_{\rm BHZ}$ along the $x$ direction. The eigenstates are normalized by
\begin{equation}
\langle\!\langle u_{m} \left( k_{x} \right) | u_{n} \left( k'_{x} \right) \rangle
= \langle u_{m} \left( k_{x} \right) | u_{n} \left( k'_{x} \right) \rangle\!\rangle
= \delta_{m, n} \delta_{k_{x}, k'_{x}}.
\end{equation}
Then, a right (left) eigenstate of $H_{\rm BHZ}$ is given by $\ket{k_{x}} \ket{u_{n} \left( k_{x} \right)}$ ($\ket{k_{x}} | u_{n} \left( k_{x} \right) \rangle\!\rangle$) with
\begin{equation}
\ket{k_{x}} := \frac{1}{\sqrt{L_{x}}} \sum_{x=1}^{L_{x}} e^{\ii xk_{x}} \ket{x},~
k_{x} \in \left\{ 0, \frac{2\pi}{L_{x}}, \cdots, \frac{2 \left( L_{x} - 1\right) \pi}{L_{x}}\right\}.
\end{equation}
Using these eigenstates, we expand the initial state $\ket{\psi \left( 0 \right)} := \sum_{x, y} c_{xy} \ket{x}\ket{y}$ as
\begin{equation}
\ket{\psi \left( 0 \right)}
= \sum_{k_{x}, n} c_{n} \left( k_{x} \right) \ket{k_{x}} \ket{u_{n} \left( k_{x} \right)},
\end{equation}
with 
\begin{equation}
c_{n} \left( k_{x} \right) := \frac{1}{\sqrt{L_{x}}} \sum_{x, y} c_{xy} e^{-\ii k_{x}x} \langle\!\langle u_{n} \left( k_{x} \right) | y \rangle.
\end{equation}
This state evolves into
\begin{eqnarray}
\ket{\psi \left( t \right)}
&=& e^{-\ii H_{\rm BHZ} t} \ket{\psi \left( 0 \right)} \nonumber\\
&=& \sum_{k_{x}, n} c_{n} \left( k_{x} \right) e^{-\ii E_{n} \left( k_{x} \right) t} \ket{k_{x}} \ket{u_{n} \left( k_{x} \right)},
\end{eqnarray}
and its amplitude at $y=y_{0}$ is 
\begin{equation}
\left| \braket{y_{0} | \psi \left( t \right)} \right|^{2}
= \sum_{k_{x}} \left| \sum_{n} c_{n} \left( k_{x} \right) \braket{y_{0} | u_{n} \left( k_{x} \right)} e^{-\ii E_{n} \left( k_{x} \right) t} \right|^{2}.
\end{equation}

Figure~\ref{fig: power oscillation} shows the evolutions of the population at the edge $y_{0} = 1$ for each phase. There, an initial state is prepared to be a localized state at the edge $y_{0} =1$. In the topological phase, the wavepacket remains localized because of the presence of helical edge states, while some of the population is absorbed into the bulk. The helical edge states indeed exhibit oscillatory dynamics. Although the edge amplitude oscillates even in the Hermitian case, the oscillation is enhanced by non-Hermiticity and the consequent nonorthogonality. In the trivial phase, on the other hand, the wavepacket quickly diffuses into the bulk since no edge states appear, which results in the monotonic decrease in the edge amplitudes in both Hermitian and non-Hermitian cases. Such power oscillation of the nonorthogonal edge states can in principle occur even in non-Hermitian topological systems with complex spectra. However, it is in practice difficult to observe because amplification or attenuation dominates the nonunitary dynamics and clears away a signature of the power oscillation.

\section{Discussion}
	\label{sec: conclusion}

The reality of spectra is relevant to the stability of non-Hermitian systems. Nevertheless, non-Hermiticity often makes spectra of bulk or edges in topological insulators complex. In this work, we have shown that a combination of pseudo-Hermiticity and reciprocity (a variant of time-reversal symmetry) enables the entirely real spectra even in non-Hermitian topological insulators. Thanks to pseudo-Hermiticity, the bulk spectra remain real as long as an energy gap for the real part of the spectrum is open. Still, the gapless edge states are not necessarily real solely in the presence of pseudo-Hermiticity. Instead, the reality of the edge spectrum is ensured by Kramers degeneracy due to reciprocity. As a prototypical example, we have illustrated this with a non-Hermitian extension of the BHZ model~\cite{BHZ-06}. Although Ref.~\cite{Hu-11} showed that the entirely real spectra of both bulk and edges are impossible in a large class of non-Hermitian topological insulators with parity-time symmetry, the discussion there is not directly applicable in the presence of additional symmetry such as pseudo-Hermiticity and reciprocity.

Non-Hermitian topological insulators with real spectra can be experimentally realized in various synthetic materials. In fact, Hermitian $\mathbb{Z}_{2}$ topological insulators including the BHZ model can be created in a variety of classical systems, such as photonic systems~\cite{Hafezi-11, *Hafezi-13, Khanikaev-13}, mechanical metamaterials~\cite{Susstrunk-15}, and electric circuits~\cite{Lee-18}. In these systems, non-Hermiticity such as gain or loss, as well as asymmetric hopping, can be introduced by judiciously controlling the external coupling to the environment~\cite{Konotop-review, Christodoulides-review}. An experimental signature of the entirely real spectra is the power oscillation of helical edge states, which is induced by the nonorthogonality due to non-Hermiticity.

Moreover, real spectra may be feasible in non-Hermitian topological insulators with different symmetry in different spatial dimensions. As long as internal symmetry is relevant, they can be systematically explored on the basis of topological classification of non-Hermitian systems~\cite{KSUS-19}. Spatial symmetry can also enrich band structures of non-Hermitian systems. Furthermore, a recent work demonstrated the entirely real spectrum in a non-Hermitian topological quasicrystal in one dimension~\cite{Zeng-20}. Further research is warranted for such new types of non-Hermitian topological insulators with real spectra.

\section*{Acknowledgment}
We thank Dmitry V. Savin and Qi-Bo Zeng for bringing our attention to Ref.~\cite{Savin-97} and Refs.~\cite{Zeng-16, Zeng-20}, respectively. This work was supported by a Grant-in-Aid for Scientific Research on Innovative Areas ``Topological Materials Science" (KAKENHI Grant No.~JP15H05855) from the Japan Society for the Promotion of Science (JSPS), and JST CREST Grant No.~JPMJCR19T2. K.K. was supported by KAKENHI Grant No.~JP19J21927 from the JSPS. M.S. was supported by KAKENHI Grant No.~JP17H02922 and No.~JP20H00131 from the JSPS.

\appendix

\section{Non-Hermitian Bernevig-Hughes-Zhang model protected by time-reversal symmetry}
	\label{appendix: BHZ-complex}

In Sec.~\ref{sec: BHZ}, we have investigated a non-Hermitian extension of the BHZ model protected by reciprocity. While reciprocity is equivalent to time-reversal symmetry in Hermitian systems, this is not the case in non-Hermitian systems. In fact, time-reversal symmetry in non-Hermitian spinful systems is defined by Eq.~(\ref{eq: TRS-cc}), which is different from reciprocity in Eq.~(\ref{eq: TRS}). Both symmetry can protect the topological phase of the BHZ model as long as the real part of the spectrum is gapped. However, the real spectrum of the helical edge states cannot be protected by time-reversal symmetry, which contrasts with reciprocity.

To see the difference between reciprocity and time-reversal symmetry, we here consider another non-Hermitian extension of the BHZ model protected by time-reversal symmetry:
\begin{eqnarray}
&&\tilde{H}_{\rm BHZ} \left( {\bm k} \right)
= \left( m + t \cos k_{x} + t \cos k_{y} \right) \tau_{z} + t \left( \sin k_{y} \right) \tau_{y} \nonumber \\
&&\qquad\qquad\qquad\quad + t \left( \sin k_{x} \right) \sigma_{z} \tau_{x} + \ii \gamma \sigma_{x} \tau_{x}.
	\label{eq: NH-BHZ2}
\end{eqnarray}
In a similar manner to the previous model $H_{\rm BHZ} \left( {\bm k} \right)$, this model $\tilde{H}_{\rm BHZ} \left( {\bm k} \right)$ respects pseudo-Hermiticity in Eq.~(\ref{eq: pH}) with $\eta = \sigma_{z}$:
\begin{equation}
\sigma_{z} \tilde{H}_{\rm BHZ}^{\dag} \left( {\bm k} \right) \sigma_{z} = \tilde{H}_{\rm BHZ} \left( {\bm k} \right).
\end{equation} 
By contrast, it does not respect reciprocity in Eq.~(\ref{eq: TRS}); instead, it respects time-reversal symmetry in Eq.~(\ref{eq: TRS-cc}) with $\mathcal{T} = \ii\sigma_{y}$:
\begin{equation}
\left( \ii \sigma_{y} \right) \tilde{H}_{\rm BHZ}^{*} \left( {\bm k} \right) \left( \ii \sigma_{y} \right)^{-1} = \tilde{H}_{\rm BHZ} \left( -{\bm k} \right).
\end{equation} 
Notably, a similar non-Hermitian quantum spin Hall insulator was also investigated in Ref.~\cite{KHGAU-19}.

\begin{figure}[H]
\centering
\includegraphics[width=86mm]{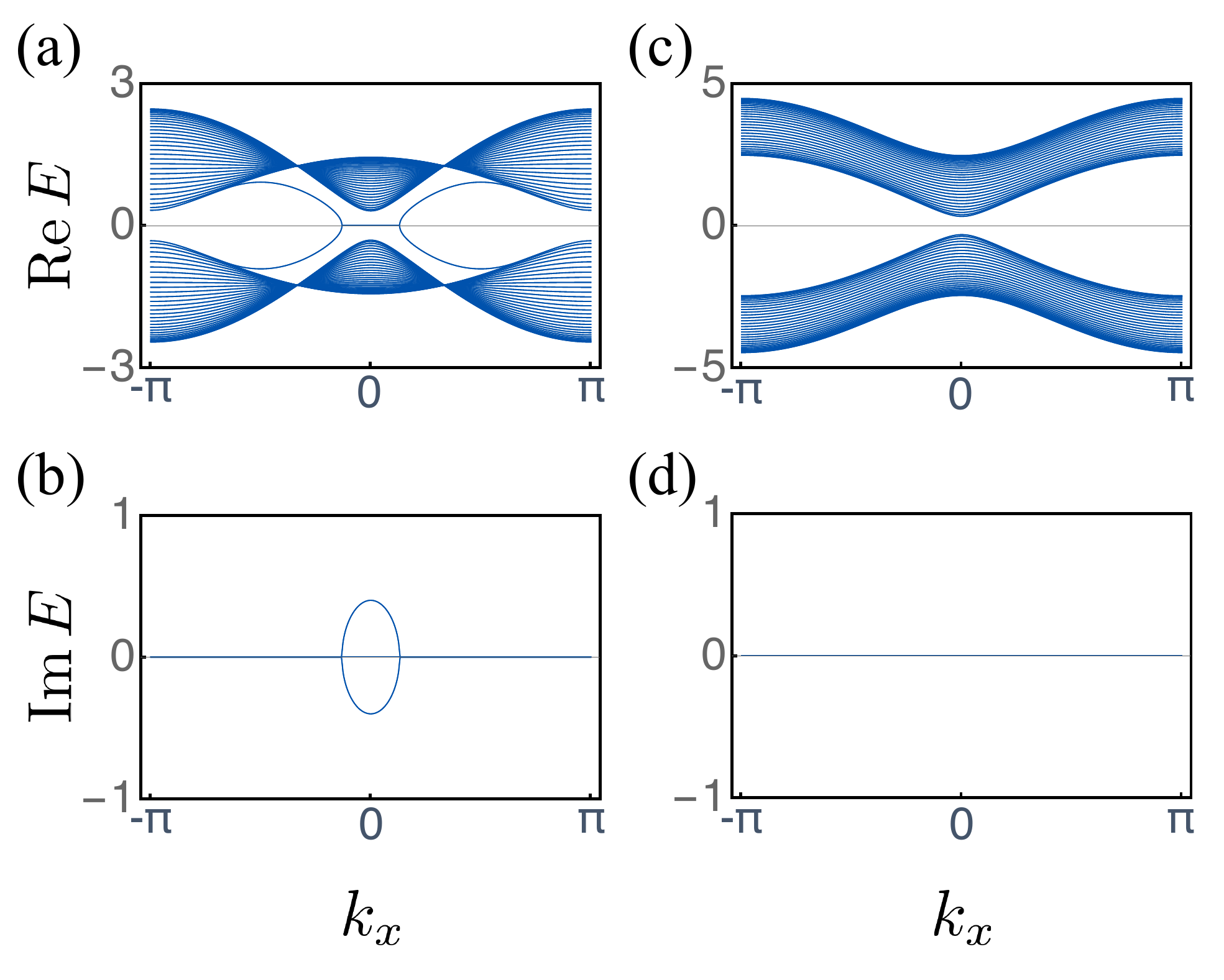} 
\caption{Complex spectrum of the non-Hermitian Bernevig-Hughes-Zhang model protected by time-reversal symmetry. The open boundary conditions are imposed in the $y$ direction ($30$ sites), whereas the periodic boundary conditions are imposed in the $x$ direction, along which the wavenumber $k_{x}$ is defined. (a, b)~Gapped and topologically nontrivial phase ($t = 1.0$, $m = -0.5$, $\gamma = 0.4$). A pair of helical edge states appears around $k_{x} = 0$. The helical edge states coalesce with each other and form exceptional points, leading to the complex spectrum at the edges. (c, d)~Gapped and topologically trivial phase ($t = 1.0$, $m = -2.5$, $\gamma = 0.4$). No edge states appear between the gapped bands, and the spectrum is entirely real.}
	\label{fig: spectrum-complex}
\end{figure}

The spectrum of $\tilde{H}_{\rm BHZ} \left( {\bm k} \right)$ is shown in Fig.~\ref{fig: spectrum-complex}. The bulk spectrum is real as long as the bulk bands are gapped, which is due to pseudo-Hermiticity. Between the gapped bulk bands, a pair of helical edge states appears in the topological phase. In the previous model $H_{\rm BHZ} \left( {\bm k} \right)$, these helical edge states are forbidden to mix with each other because of the Kramers degeneracy. However, time-reversal symmetry does not impose such a constraint in non-Hermitian systems. Consequently, the helical edge states coalesce with each other and form a pair of exceptional points in the present model $\tilde{H}_{\rm BHZ} \left( {\bm k} \right)$; the edge spectrum becomes complex. Physically, the complex edge spectrum means the amplification (lasing) of the helical edge states.

\bibliography{NH-topo}

\end{document}